\documentclass[journal=jacsat,manuscript=article]{achemso}

\usepackage{chemformula} % Formula subscripts using \ch{}
\usepackage[T1]{fontenc} % Use modern font encodings

\author{Viswesh Prakash}
\author{Jingyu Tang}
\author{Lisa M. Porter}

\author{Rachel C. Kurchin}
\email{rkurchin@cmu.edu}

\affiliation[Carnegie Mellon University]
{Department of Materials Science and Engineering, Carnegie Mellon University, Pittsburgh, Pennsylvania 15213, USA}

\title[]
  {First-Principles Study of Mg-Induced Phase Stabilization in \ch{Ga2O3} polymorphs} %%\footnote{A footnote for the title}}

\begin{document}

%%%%%%%%%%%%%%%%%%%%%%%%%%%%%%%%%%%%%%%%%%%%%%%%%%%%%%%%%%%%%%%%%%%%%
%% The "tocentry" environment can be used to create an entry for the
%% graphical table of contents. It is given here as some journals
%% require that it is printed as part of the abstract page. It will
%% be automatically moved as appropriate.
%%%%%%%%%%%%%%%%%%%%%%%%%%%%%%%%%%%%%%%%%%%%%%%%%%%%%%%%%%%%%%%%%%%%%
\begin{tocentry}

% Some journals require a graphical entry for the Table of Contents.
% This should be laid out ``print ready'' so that the sizing of the
% text is correct.

% Inside the \texttt{tocentry} environment, the font used is Helvetica
% 8\,pt, as required by \emph{Journal of the American Chemical
% Society}.

% The surrounding frame is 9\,cm by 3.5\,cm, which is the maximum
% permitted for  \emph{Journal of the American Chemical Society}
% graphical table of content entries. The box will not resize if the
% content is too big: instead it will overflow the edge of the box.

% This box and the associated title will always be printed on a
% separate page at the end of the document.
\includegraphics[width=7.8cm]{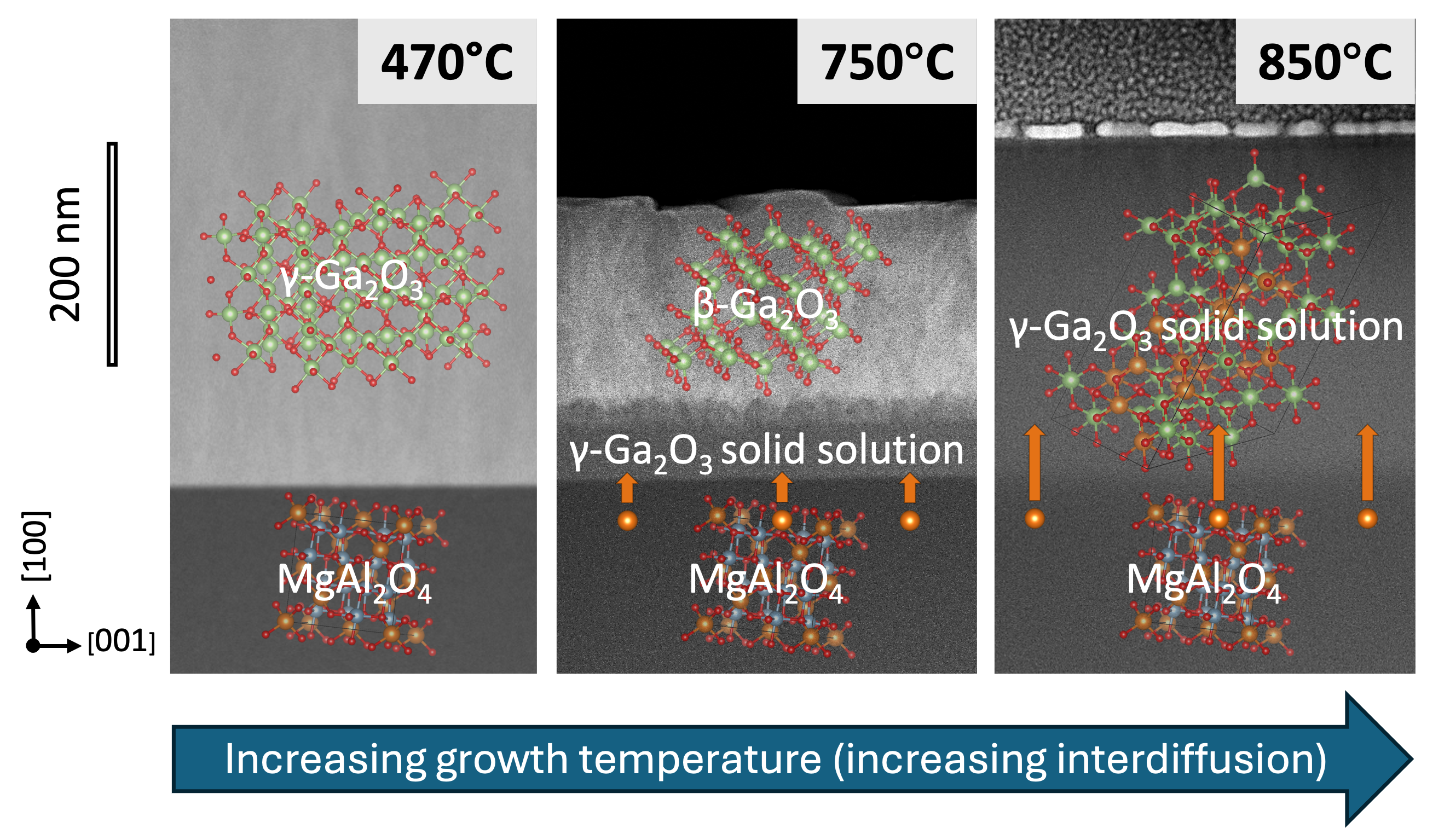}

\end{tocentry}

%%%%%%%%%%%%%%%%%%%%%%%%%%%%%%%%%%%%%%%%%%%%%%%%%%%%%%%%%%%%%%%%%%%%%
%% The abstract environment will automatically gobble the contents
%% if an abstract is not used by the target journal.
%%%%%%%%%%%%%%%%%%%%%%%%%%%%%%%%%%%%%%%%%%%%%%%%%%%%%%%%%%%%%%%%%%%%%
\begin{abstract}
  In this study, we investigate the effect of Mg incorporation on the relative phase stability of the four primary \ch{Ga2O3} polymorphs using density functional theory (DFT) calculations, with the goal of rationalizing experimental observations suggesting that diffusion from \ch{MgAl2O4} substrates contributes to relative stabilization of the $\gamma$ phase. Mg incorporation is modeled up to 25\% of Ga sites within supercells derived from fully relaxed unit cells of each polymorph. %We outline and employ a statistical approach to select representative subsets of alloyed structures based on radial distribution functions and the Kullback–Leibler divergence metric for systems with a large configurational space. 
  Our results show that while $\beta$-\ch{Ga2O3} remains the thermodynamically most stable phase, the enthalpic differences between polymorphs decrease with increasing Mg content. The inherently disordered $\gamma$ phase, with its high configurational entropy, becomes less energetically unfavorable under Mg substitution, suggesting that entropy-driven stabilization may facilitate its formation under high-temperature and/or nonequilibrium growth conditions such as those previously reported. These findings provide a thermodynamic rationale for the experimental observation of the $\gamma$ phase during epitaxial growth on \ch{MgAl2O4} spinel substrates.
\end{abstract}

%%%%%%%%%%%%%%%%%%%%%%%%%%%%%%%%%%%%%%%%%%%%%%%%%%%%%%%%%%%%%%%%%%%%%
%% Start the main part of the manuscript here.
%%%%%%%%%%%%%%%%%%%%%%%%%%%%%%%%%%%%%%%%%%%%%%%%%%%%%%%%%%%%%%%%%%%%%
\section{Introduction}
Gallium oxide (\ch{Ga2O3}) has emerged as a material of considerable interest for high-power electronic applications, owing to its ultra-wide bandgap, high breakdown field, and potential for efficient power device performance.\cite{higashiwaki,green2017,jadhav,liu2022} \ch{Ga2O3} has four widely accepted polymorphs showing the following symmetries: $\alpha$ (trigonal), $\beta$ (monoclinic), $\gamma$ (cubic), and $\kappa$ (also referred to as $\epsilon$, orthorhombic). Among these, $\beta$-\ch{Ga2O3} is the thermodynamically stable phase,\cite{tang2024thermal} while $\gamma$-\ch{Ga2O3} is consistently identified as the least stable based on density functional theory (DFT) calculations.\cite{yoshioka2007,playford_structures_2013}

Despite its theoretical instability, the $\gamma$ phase is frequently observed in experimental studies. Notably, the $\gamma$-\ch{Ga2O3} phase has been reported to form as inclusions during the epitaxial growth of $\beta$-\ch{Ga2O3} films on $\beta$-\ch{Ga2O3} substrates, particularly when alloyed with \ch{Al2O3} or following ion implantation.\cite{chang2021,garcia2022} Additionally, thin $\gamma$-\ch{Ga2O3} layers—typically on the order of a few tens of nanometers—have been detected at the interfaces of $\beta$-\ch{Ga2O3} films grown on \ch{MgO} substrates with various crystallographic orientations, including (001), (011), and (111).\cite{vura2022,nakagomi2017} These observations suggest that specific growth conditions and substrate interactions can kinetically favor the formation of the metastable $\gamma$ phase.

\begin{figure}
    \centering
    \includegraphics[width=1\linewidth]{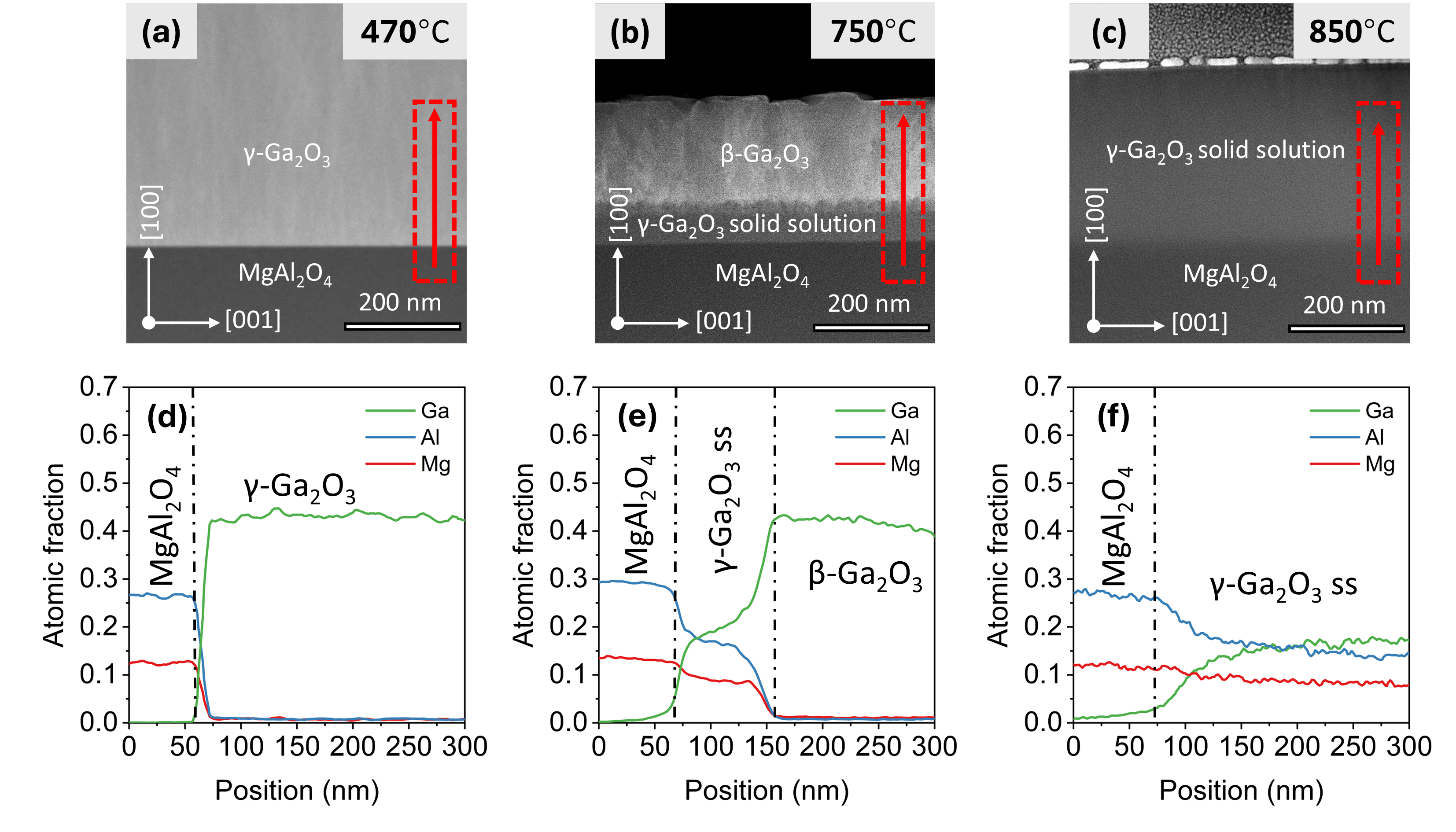}
    \caption{Cross-sectional high-angle annular dark-field scanning transmission electron microscopy (HAADF-STEM) images of films grown at (a) 470 \textdegree C, (b) 750 \textdegree C, and (c) 850 \textdegree C, respectively and their corresponding integrated EDX composition profiles shown in (d-f). The red dashed rectangles with arrows indicate the regions and directions along which the EDX profiles were acquired.}
    \label{fig:stem_edx}
\end{figure}

Experimental evidence also indicates that the incorporation of certain elements during film growth can contribute to the stabilization of the $\gamma$ phase. In particular, the addition of Mg\cite{hou2022,hou_mocvd_2022,jingyumg}, Al\cite{bhuiyan2020}, Mn\cite{huang_microstructure_2007,huang_effect_2020,hayashi2006}, Fe\cite{huang2020}, or Cu\cite{liu_stabilizing_2018} has been found to promote the formation or retention of the $\gamma$ phase. Our previous studies revealed the sequence of $\gamma$-\ch{Ga2O3} $\rightarrow$ $\beta$-\ch{Ga2O3} $\rightarrow$ $\gamma$-\ch{Ga2O3} solid solution on (100)-oriented \ch{MgAl2O4} substrates grown by metal-organic chemical vapor deposition (MOCVD) as growth temperature is increased. Specifically, $\gamma$-\ch{Ga2O3} formed at 470 \textdegree C, followed by the emergence of $\beta$-\ch{Ga2O3} in the temperature range of 530–650 \textdegree C. At higher growth or annealing temperatures (>750 \textdegree C), substantial Mg and Al diffusion occurred, leading to the formation of a $\gamma$-\ch{Ga2O3} solid-solution.\cite{tang2024atomic, jingyumg, jiang2023evolution}  Scanning transmission electron microscopy (STEM) analyses conducted across the substrate/thin-film interfaces of these films are shown in Figure~\ref{fig:stem_edx}. %revealed evidence of significant chemical interdiffusion—specifically, the migration of Mg and Al from the substrate that contributed to the stabilization of the $\gamma$ phase.
In this figure, it can be observed that at the intermediate growth temperature (750 \textdegree C), $\beta$-\ch{Ga2O3} becomes the preferred phase in the upper portion of the film where Mg and Al incorporation is negligible, while a transition layer below with substantial Mg and Al incorporation exhibits the $\gamma$ structure. However, by 850 \textdegree C, diffusion is fast enough that this $\beta$ layer is no longer present, and $\gamma$ phase solid solution is stabilized throughout the film.
More detailed growth conditions and structural characterizations of these films can be found in our previous publications.\cite{jingyumg,tang2024atomic, jiang2023evolution} 

 %The STEM energy dispersive X-ray spectroscopy (EDX) analysis in Figure ~\ref{fig:stem_edx} reveals the sequence of $\gamma$-\ch{Ga2O3} $\rightarrow$ $\beta$-\ch{Ga2O3} $\rightarrow$ $\gamma$-\ch{Ga2O3} solid solution, driven by Mg and Al interdiffusion from the substrate. As shown in Figures~\ref{fig:stem_edx}(a) and~\ref{fig:stem_edx}(d), in the absence of Mg and Al interdiffusion, $\gamma$-\ch{Ga2O3} remains the least stable phase of $\ch{Ga2O3}$ and can only be grown within a narrow growth window and much lower growth temperatures than other phases such as $\alpha$-\ch{Ga2O3} or $\kappa$-\ch{Ga2O3} in MOCVD. At 750 \textdegree C (Figures~\ref{fig:stem_edx}(b) and~\ref{fig:stem_edx}(e)), 
 
 %Previous studies have demonstrated that $\beta$-\ch{Ga2O3} nucleates first under the similar growth condition, followed by the Mg and Al interdiffusion.\cite{jiang2023evolution} At 850 \textdegree C (Figures~\ref{fig:stem_edx}(c) and~\ref{fig:stem_edx}(f)), Mg and Al diffuse throughout the entire film thickness, resulting in the formation of a $\gamma$-structured \ch{Ga2O3} solid solution. 

Further evidence for the role of \ch{MgAl2O4} substrates in stabilizing the $\gamma$ phase has been provided by Oshima et al.\cite{oshima_epitaxial_2012} That study reports X-ray diffraction peak shifts indicative of diffusion-stabilized $\gamma$-\ch{Ga2O3} formation in films grown via mist chemical vapor deposition (mist-CVD) on (100)-oriented \ch{MgAl2O4}. Hou et al.\cite{hou2022,hou_mocvd_2022} also reported the successful growth of (111)-oriented, phase-pure Mg\textsubscript{x}Ga\textsubscript{2}O\textsubscript{4} inverse spinel films on (0001) sapphire substrates, without the formation of the $\beta$ phase. Furthermore, they observed that post-growth annealing at 900 \textdegree C led to improved structural ordering without inducing a phase transformation to the $\beta$ structure.

In addition to experimental demonstrations,\cite{bhuiyan2020,watanabe_synthesis_2011} theoretical analyses indicate that the $\gamma$ phase can become the dominant structural form in (Al\textsubscript{x}Ga\textsubscript{1-x})\textsubscript{2}O\textsubscript{3} films depending on the Al concentration and specific growth technique employed. Several computational studies have sought to elucidate the effects of alloying on the phase stability of \ch{Ga2O3} polymorphs. Peelaers et al.\cite{peelaers_structural_2018} investigated the relative stability of monoclinic ($\beta$)  and corundum ($\alpha$) (Al\textsubscript{x}Ga\textsubscript{1-x})\textsubscript{2}O\textsubscript{3} alloys at zero temperature, while Seacat et al.\cite{seacat_orthorhombic_2020} extended this analysis to the orthorhombic ($\kappa$) phase. Mu and Van de Walle\cite{mu_phase_2022} conducted a comprehensive analysis of Al incorporation into all known \ch{Ga2O3} polymorphs, demonstrating that alloying with Al significantly reduces the energy difference between the $\gamma$- and $\beta$ phases, thereby enhancing the metastability of $\gamma$-\ch{Ga2O3}.

While the stabilizing effect of Al alloying on the $\gamma$ phase has been explored in detail, experimental observations suggest that both Al and Mg play critical roles in stabilizing the $\gamma$ phase under specific growth conditions. However, the independent influence of Mg incorporation on the phase stability of \ch{Ga2O3} polymorphs remains an open question and is largely unaddressed in the existing literature.

The present study seeks to address this gap by employing a first-principles computational approach based on density functional theory. We investigate the relative phase stability of all four \ch{Ga2O3} polymorphs—$\alpha$, $\beta$, $\gamma$, and $\kappa$—in (Mg\textsubscript{x}Ga\textsubscript{1-x})\textsubscript{2}O\textsubscript{3-x} alloys at zero temperature. Through this work, we aim to clarify the thermodynamic role of Mg incorporation in stabilizing the metastable $\gamma$ phase and to contribute a more complete understanding of the compositional factors that govern polymorph formation in \ch{Ga2O3}-based materials.

\section{Methodology}

\subsection{Structures of Polymorphs}

In this study, we focus on the four commonly accepted polymorphs of \ch{Ga2O3}—$\beta$ (monoclinic), $\kappa$ (orthorhombic), $\alpha$ (trigonal), and $\gamma$ (cubic)—as well as the \ch{MgO} structure, which adopts a cubic rock-salt configuration. Among these \ch{Ga2O3} polymorphs, the $\gamma$ phase is characterized by a disordered structure\cite{playford_structures_2013,playford_characterization_2014}, where cation vacancies can occupy both tetrahedral and octahedral sites in the defective spinel structure. Hence the unit cell of $\gamma$-\ch{Ga2O3} was constructed following the two-step process outlined in the literature \cite{yoshioka2007,mu_phase_2022,gutierrez_theoretical_2001}. First, a triclinic supercell was derived from the conventional cubic spinel cell corresponding to the stoichiometry  \ch{Ga3O4} by applying a lattice transformation to the lattice vectors $(\vec a_0, \vec b_0, \vec c_0)$ as follows:
\begin{equation} \label{eq1}
\begin{pmatrix}
\vec{a'}\\
\vec{b'}\\
\vec{c'}\\
\end{pmatrix}
=
\begin{pmatrix}
0.5 & 0.5 & 0\\
0 & 0.5 & 0.5\\
1.5 & 0 & 1.5\\
\end{pmatrix}
\begin{pmatrix}
\vec a_0\\
\vec b_0\\
\vec c_0\\
\end{pmatrix}
\end{equation}

\begin{figure}
    \centering
    \includegraphics[width=1\linewidth ]{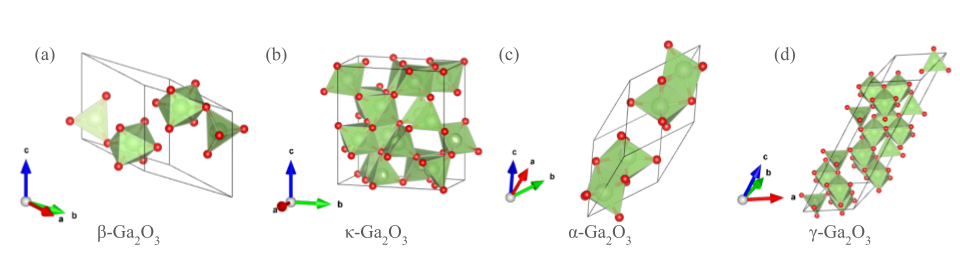}
    \caption{Illustration of structures for \ch{Ga2O3} polymorphs in decreasing order of stability as determined by DFT – (a) $\beta$-\ch{Ga2O3}; (b) $\kappa$-\ch{Ga2O3}; (c) $\alpha$-\ch{Ga2O3}; (d) $\gamma$-\ch{Ga2O3}. Red spheres denote O atoms and green spheres and polyhedra denote Ga atoms and their coordination. Cells (a) and (c) are primitive unit cells of the respective structures. Structural visualization was performed using VESTA. \cite{momma_vesta_2011}}
    \label{fig:polymorph_structures}
\end{figure}

Subsequently, two Ga atoms were selectively removed from octahedral sites in the transformed triclinic cell, thereby yielding the desired stoichiometry of \ch{Ga2O3}. These Ga atoms were chosen such that they are separated by a distance of approximately 7.9 \AA.
The unit cells for all the \ch{Ga2O3} polymorphs are illustrated in Figure~\ref{fig:polymorph_structures}. The types of cation and anion sites characteristic of each polymorph are summarized in Table~\ref{tbl:1}.

\begin{table}
  \caption{Summary of Ga coordination and O site connectivity for structures of all \ch{Ga2O3} polymorphs.}
  \label{tbl:1}
  \begin{tabular}{lll}
    \hline
    Structure & Ga site ratio & Types of O sites\\
     & (octahedral:tetrahedral) & [ O at corners of polyhedra ] \\
    \hline
    $\beta$   & 1:1 & Edge sharing between octahedra  \\
     & & Corner sharing between tetrahedra \\
     & & Corner sharing between octahedra and tetrahedra \\
     \\
    $\kappa$ & 3:1 &Edge sharing between octahedra \\
    & & Corner sharing between tetrahedra \\
    & & Corner sharing between octahedra and tetrahedra \\
    \\
     $\alpha$ & All octahedral &Edge sharing between octahedra \\
    \\
    $\gamma$ & 5:3 &Edge sharing between octahedra \\
    & & Corner sharing between tetrahedra \\
    & & Corner sharing between octahedra and tetrahedra \\
    \\
    \hline
  \end{tabular}
\end{table}

To examine the effects of Mg incorporation, we construct alloyed structures by substituting Ga atoms with Mg within supercells derived from the respective unit cells shown in Figure~\ref{fig:polymorph_structures}. At each specified Mg concentration, various configurations can be generated by selectively replacing Ga atoms occupying distinct cation sites. To preserve charge balance, oxygen vacancies are introduced alongside Mg substitution (see Computational Details for more information). In this study, we restrict the extent of Mg substitution to a maximum of 25\% of the available Ga sites.

\subsection{Computational Details}

All first-principles calculations were performed using DFT within the projector augmented-wave\cite{blochl_projector_1994} (PAW) formalism as implemented in the Quantum ESPRESSO package. The exchange-correlation interactions were described using the generalized gradient approximation (GGA) as formulated by Perdew, Burke, and Ernzerhof (PBE)\cite{perdew_generalized_1996}.
Prior to structural relaxations, convergence tests were carried out to determine appropriate values for the plane-wave kinetic energy cutoff, the charge density kinetic energy cutoff, and the Brillouin-zone sampling mesh. Based on these tests, the plane-wave energy cutoff was fixed at 140 Ry, and the charge density cutoff was set to 560 Ry. Brillouin-zone integrations were performed using a $\Gamma$-centered Monkhorst–Pack 6×6×6 k-point mesh for the primitive cells of the monoclinic $\beta$ and trigonal $\alpha$ phases, a 6×4×4 mesh for the orthorhombic $\kappa$ phase, and a 6×6×2 mesh for the triclinic cell of the defective spinel $\gamma$ phase. All polymorphs were fully relaxed with respect to both atomic positions and lattice parameters. The resulting optimized structural parameters are summarized in Table~\ref{tbl:2} and are in good agreement with experimentally reported values.

\begin{table}
  \caption{DFT-relaxed structural parameters for all polymorphs of \ch{Ga2O3} in comparison to experimentally determined lattice parameters. Lattice parameters in Å and angles in degrees.}
  \label{tbl:2}
  \begin{tabular}{lll}
    \hline
    & Experimental & Calculated\\
    \hline
    Mononclinic ($\beta$) \cite{higashiwaki_-ga2o3_2022}   & &  \\
     a& 12.2 & 12.4565 \\
     b& 3.0 & 3.085 \\
     c& 5.8 & 5.879 \\
     beta& 104 & 103.688 \\
     \\
    Trigonal ($\alpha$) \cite{marezio_bond_1967} &  & \\
    a& 4.98& 5.0597 \\
    c& 13.43& 13.6283\\
    \\
    Orthorhombic ($\kappa$) \cite{cora_real_2017} &  & \\
    a& 5.0& 5.1243 \\
    b& 8.68& 8.8023 \\
    c& 9.23& 9.4124\\
    \\
    Defective spinel ($\gamma$) \cite{playford_characterization_2014} & & \\
    a& 8.24 & 8.325 \\
    \hline
  \end{tabular}
\end{table}

To minimize the effects of short-range ordering induced by periodic boundary conditions in DFT calculations of substitutional alloys, supercells were constructed for all polymorphs based on the respective unit cells shown in Figure ~\ref{fig:polymorph_structures}. Furthermore, to avoid ordering along specific lattice directions, supercells were constructed using lattice transformation matrices that yield approximately cubic supercells with a comparable distribution of Ga sites along each axis of the supercell. These lattice transformations are carried out as follows:

\begin{equation} \label{eq2}
\begin{pmatrix}
\vec{a'}\\
\vec{b'}\\
\vec{c'}\\
\end{pmatrix}
=
A_{\text{sc}}
\begin{pmatrix}
\vec{a}\\
\vec{b}\\
\vec{c}\\
\end{pmatrix}
\end{equation}

Where $\vec{a}$, $\vec{b}$, and $\vec{c}$ are the lattice vectors of the primitive cell, $A_{\text{sc}}$ is the integer lattice transformation matrix, and $\vec{a'}$, $\vec{b'}$, and $\vec{c'}$ are the resulting lattice vectors of the generated supercell, expressed as linear combinations of the original lattice vectors.

The lattice transformation matrices used for the $\beta$, $\alpha$, and $\kappa$ phases are as follows:

\begin{gather*}
A_{\beta}
=
\begin{pmatrix}
2 & 0 & 0\\
-4 & 4 & 0\\
0 & 0 & 2\\
\end{pmatrix}
\end{gather*}

\begin{align*}
A_{\alpha}
=
\begin{pmatrix}
2 & 0 & 0\\
-1 & 3 & -1\\
-1 & -1 & 3\\
\end{pmatrix} 
\end{align*}

\begin{align*}
A_{\kappa}
=
\begin{pmatrix}
2 & 0 & 0\\
0 & 1 & 0\\
0 & 0 & 1\\
\end{pmatrix} 
\end{align*}

For the defective spinel $\gamma$-\ch{Ga2O3} having a primitive cell showing triclinic symmetry, we adopt the method of constructing an oriented bulk structure as described by  Sun and Ceder\cite{sun_efficient_2013}, to ensure that the out-of-plane lattice vector aligns with experimentally observed orientations. The resulting transformation matrix from this approach is:

\begin{align*}
A_{\gamma}
=
\begin{pmatrix}
0 & 2 & 0\\
-3 & 0 & 1\\
2 & -1 & 0\\
\end{pmatrix} 
\end{align*}

Applying the aforementioned lattice transformations to the unit cells results in supercells containing 160 atoms for the $\beta$, $\alpha$, and $\gamma$ polymorphs, and 80 atoms for the $\kappa$ polymorph. These supercells serve as the basis for modeling Mg incorporation into the \ch{Ga2O3} lattice by substituting Ga atoms with Mg.
Since $\mathrm{Mg}^{2+}$ has a lower valence state than $\mathrm{Ga}^{3+}$, charge compensation is required to maintain overall neutrality. This is achieved by introducing one oxygen vacancy for every two Ga sites substituted with Mg. Thus, the alloyed compositions studied in this work follow the general formula $(\mathrm{Mg}_{x}\mathrm{Ga}_{1-x})_2\mathrm{O}_{3-x}$.

Mg substitution was explored both with and without cation site preference, i.e. Mg atoms are substituted at Ga sites both at random and by restricting to Ga atoms occupying a specific cation site (octahedral or tetrahedral). However, given the combinatorially large number of possible configurations—for instance, at 25\% of Ga site having Mg substitution, there are over $10^{14}$ possible configurations when symmetry equivalence between arrangements is not considered—an exhaustive computational study of all configurations is impractical.
To address this, we define structural equivalence using a geometric criterion: two configurations are considered symmetrically equivalent if the set of all pairwise distances among Mg atoms, among oxygen vacancies, and between Mg and vacancy sites are identical. After this symmetry-based filtering is applied, the number of inequivalent configurations is reduced at lower concentration of Mg substitution. Nevertheless, for the range of concentrations (6.25\% to 25\%) considered in this study, the number of inequivalent configurations remains substantial. Due to the computational cost associated with performing full DFT relaxations for all inequivalent structures, it is necessary to select a representative subset of configurations at each concentration for further investigation.

To select a representative subset of configurations for a given polymorph and Mg concentration, we adopt a statistical approach based on radial distribution functions (RDFs). Initially, a large number of symmetrically inequivalent structures are generated. By computing all partial RDF's between species within the structure and comparing results from subsets to averages over the full ensemble, we can select a subset that retains acceptable representation of the full range of environments. For more details, see the Supporting Information.

The thermodynamic stability of the Mg-incorporated structures at zero temperature is evaluated through the calculation of their enthalpies of formation ($\Delta$H). This is done by referencing the total energies of the substituted structures to the most stable phases of the constituent binary oxides—namely, $\beta$-\ch{Ga2O3} and \ch{MgO}. The formation enthalpy for a given composition of the alloyed system $(\mathrm{Mg}_{x}\mathrm{Ga}_{1-x})_2\mathrm{O}_{3-x}$ is computed according to the following expression (where $E[\dots]$ refers to DFT total energy):
\begin{equation}
    \Delta H \big[(\mathrm{Mg}_{x}\mathrm{Ga}_{1-x})_2\mathrm{O}_{3-x}\big] = E\big[(\mathrm{Mg}_{x}\mathrm{Ga}_{1-x})_2\mathrm{O}_{3-x}\big] - (1 - x)E\big[\mathrm{Ga}_2\mathrm{O}_3\big] - 2xE\big[\mathrm{MgO}\big]
\end{equation}

\section{Results and Discussion}
\begin{figure}
    \centering
    \includegraphics[width=0.75\linewidth]{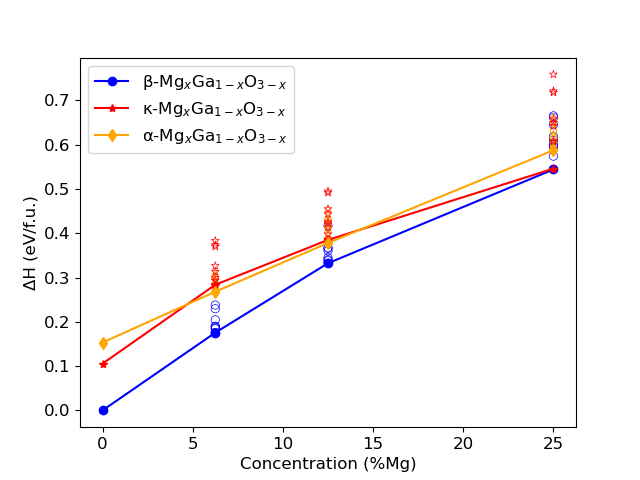}
    \caption{Bulk enthalpy of formation ($\Delta H$) for $(\mathrm{Mg}_{x}\mathrm{Ga}_{1-x})_2\mathrm{O}_{3-x}$ alloys as a function of Mg concentration (\% of Ga sites substituted) for the $\alpha$-, $\beta$-, and $\kappa$-polymorphs of \ch{Ga2O3}. Filled symbols represent minimum values, while open symbols represent the values for enthalpy of formation of other structures in the selected sample of structures for each polymorph at a given concentration.}
    \label{fig:bulk_enthalpy}
\end{figure}

The phase stability of the different \ch{Ga2O3} polymorphs was assessed by comparing their enthalpies of formation ($\Delta H$). Our results for the unsubstituted \ch{Ga2O3} structures agree with previous calculations \cite{yoshioka2007} and the relative phase stability follows the expected trend, in order of increasing $ \Delta H$: $\beta$< $\kappa$ < $\alpha$ < $\gamma$. The $\gamma$ phase, as anticipated, is the least stable, with a $ \Delta H$ of 0.222 eV/f.u.—significantly higher than those of the $\alpha$ phase ($ \Delta H$ = 0.153 eV/f.u.) and $\kappa$ phase ($ \Delta H$ = 0.106 eV/f.u.), all referenced to the $\beta$ phase. Our calculated $ \Delta H$ values are consistent with prior literature \cite{peelaers_structural_2018,seacat_orthorhombic_2020,mu_phase_2022}, deviating by  $\sim$0.01 eV/f.u. 
Figure~\ref{fig:bulk_enthalpy} presents the minimum $ \Delta H$ values for the $\beta$, $\kappa$, and $\alpha$ polymorphs as a function of Mg substitution on Ga sites, without cell shape relaxation. From Figure~\ref{fig:bulk_enthalpy}, we note that as the Mg concentration increases, the differences in enthalpy of formation across the polymorphs decrease with increasing concentration of Mg. This suggests that Mg incorporation reduces the energetic penalty for forming the metastable polymorphs, rendering them more competitive in energy with the thermodynamically stable $\beta$ phase. Consequently, the addition of Mg may make the metastable polymorphs energetically accessible structures that can be grown by tuning the growth conditions, where the influence of kinetics, strain or co-doping effects of other atoms could further stabilize the different polymorphs.

\begin{figure}
    \centering
    \includegraphics[width=1\linewidth]{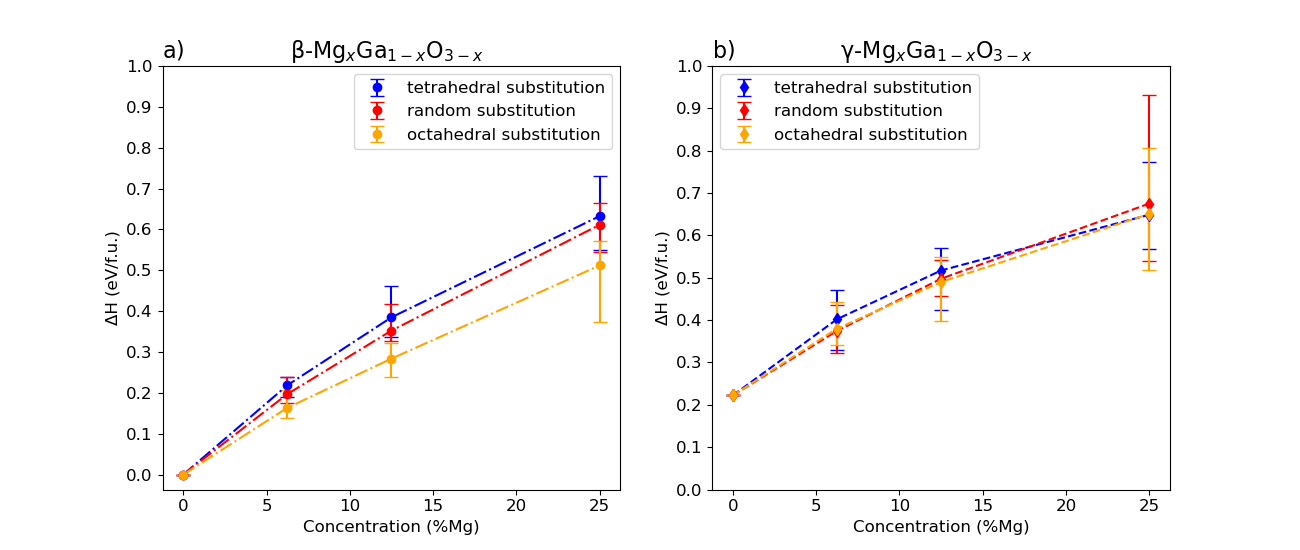}
    \caption{Average and range of enthalpy of formation ($\Delta H$) for epitaxially strained Mg-substituted (a) $\beta$-\ch{Ga2O3} and (b) $\gamma$-\ch{Ga2O3} as a function of \%  Ga sites substituted, shown for three site occupancy configurations: octahedral-only (yellow), tetrahedral-only (blue), and random distribution (red).}
    \label{fig:epi_strain}
\end{figure}

Given that both the $\gamma$- and $\beta$ phases have been experimentally observed to grow on Mg-containing substrates such as \ch{MgO} and \ch{MgAl2O4}, we examine the effect of Mg substitution in these two polymorphs in greater detail. Specifically, we analyze the site preference and energetic impact of Mg incorporation across different substitution scenarios. To ensure relevance to experimentally observed growth conditions, these calculations are performed using epitaxially constrained structures, where the out-of-plane lattice vector that was allowed to relax is consistent with the experimentally reported out-of-plane normal for epitaxial growth (and the other two lattice vectors were fixed according to the epitaxial matching constraint). Our results indicate that in $\beta$-\ch{Ga2O3}, Mg atoms show a clear preference to occupy octahedral Ga sites. Figure~\ref{fig:epi_strain}a shows the average and range of the enthalpy of formation for substituted $\beta$-\ch{Ga2O3} as a function of the Mg concentration, we see that the octahedral substitution of Mg significantly reduces the enthalpy of formation for the substituted structure at higher concentrations when compared to tetrahedral substitution or random substitution. In contrast, $\gamma$-\ch{Ga2O3} does not exhibit a strong site preference for Mg; the enthalpy of formation remains relatively insensitive to the specific site occupancy across the full range of Mg concentrations considered. This can be seen in Figure~\ref{fig:epi_strain}b, exhibiting no noticeable reduction in the enthalpy of formation for any given site occupancy of Mg in the $\gamma$ structure.

To further understand the energetic relationship between the $\gamma$- and $\beta$-polymorphs during Mg substitution, we directly compare the enthalpy of formation for both polymorphs as a function of Mg concentration for the same three three substitution configurations: octahedral-only, tetrahedral-only, and random occupancy (Figure S20). The key takeaway from this comparison is that, like in the relaxed bulk results shown in Figure~\ref{fig:bulk_enthalpy}, Mg incorporation also reduces the energy difference between the $\beta$ and $\gamma$ phases in these strained simulations.%Notably, we find that Mg substitution exclusively at tetrahedral sites leads to the greatest reduction in the enthalpy difference between the $\gamma$- and $\beta$ phases at higher concentrations, suggesting that when Mg occupies tetrahedral sites, the $\gamma$ phase becomes considerably more energetically accessible.

\begin{figure}
    \centering
    \includegraphics[width=1\linewidth]{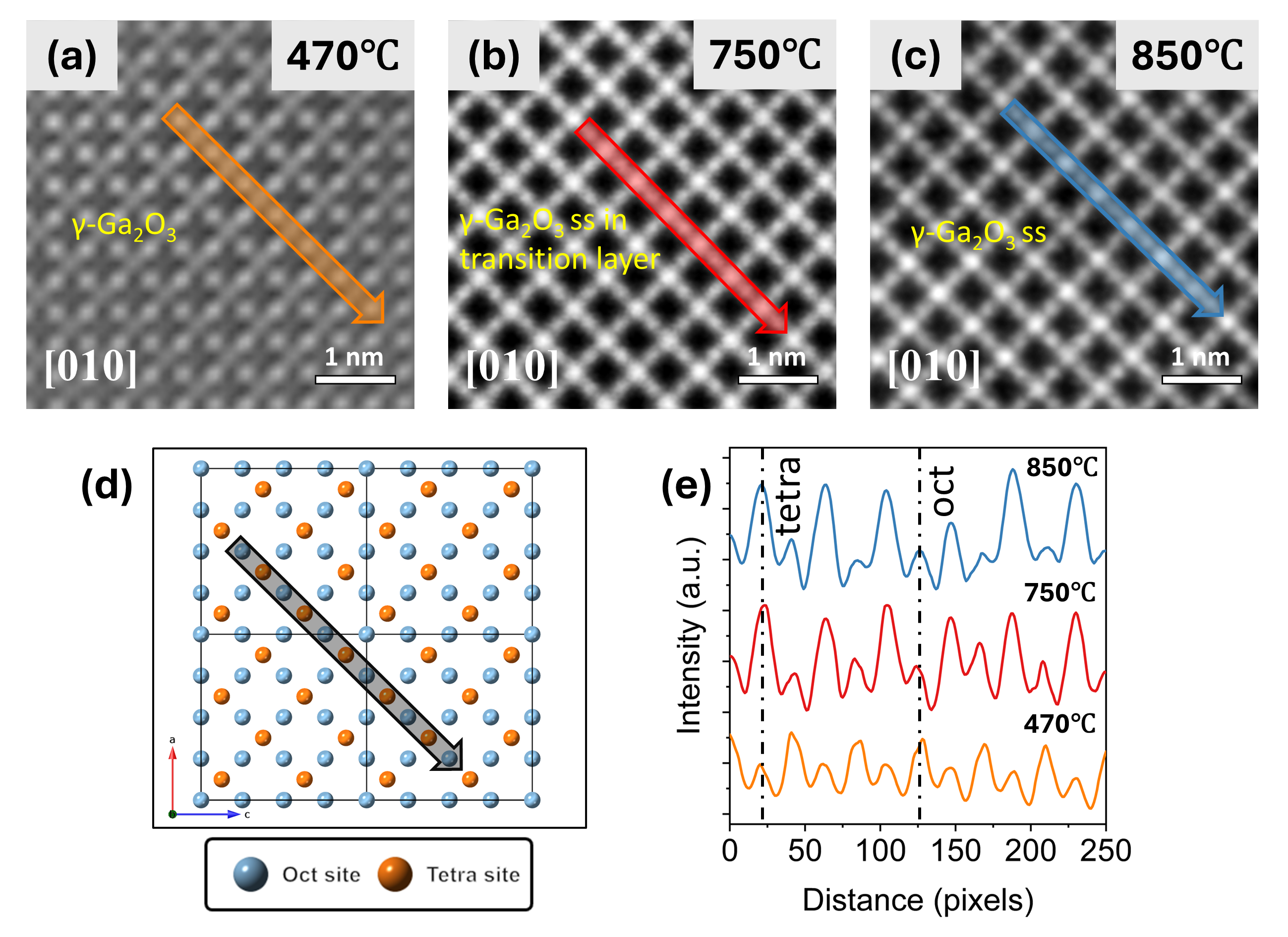}
    \caption{Atomic-scale HAADF-STEM images for the film grown at (a) 470 °C, (b) 750 °C and (c) 850 °C, respectively. (d) shows the atomic structure of cubic spinel structure along [010] zone. (e) shows the combined line profiles of the image intensity across three atomic rows from (a) to (c), showing the inversion in relative contrast between tetrahedral and octahedral sites.}
    \label{fig:stem_linescans}
\end{figure}

Figures~\ref{fig:stem_linescans}a-c show atomic-scale HAADF-STEM images of representative regions from the same three films shown in Figure~\ref{fig:stem_edx}, all exhibiting structural features consistent with the spinel structure shown in Figure~\ref{fig:stem_linescans}d. Three line profiles highlighted in orange, red and blue, each starting and ending at atoms on octahedral sites, were extracted from the images to evaluate the cation site preference based on the contrast difference between tetrahedral and octahedral sites, where the HAADF contrast is roughly $Z^{1.6}$ (where $Z$ is the atomic number). The lighter Al and Mg atoms preferentially occupy octahedral Ga sites and the heavier Ga atoms prefer to be on tetrahedral sites, as observed by the inversion of the image intensity profiles (Figure~\ref{fig:stem_linescans}e) for pure $\gamma$-\ch{Ga2O3} (Figure~\ref{fig:stem_linescans}a, orange line in e) and $\gamma$-\ch{Ga2O3} solid solutions (Figure~\ref{fig:stem_linescans}b-c, red/blue lines in e).

% In experimental studies, Mg substitution frequently occurs alongside Al incorporation, often as a result of diffusion from \ch{MgAl2O4} substrates during growth. In a study similar to the present work, Mu et al. \cite{mu_phase_2022} report that Al strongly prefers to occupy octahedral Ga sites in both $\beta$- and $\gamma$-\ch{Ga2O3}, unlike the results reported here where this preference is only present in the $\beta$ phase. This site preference causes Ga to occupy the tetrahedral sites and may, in turn, drive Mg to occupy the remaining tetrahedral sites. Taken together, our findings suggest that the co-substitution of Al and Mg may lead to a site-selective stabilization of the $\gamma$ phase, with Al occupying octahedral sites and Mg thus ``pushed'' to tetrahedral ones, which collectively lowers the thermodynamic cost of forming the metastable gamma structure. This proposed mechanism is consistent with experimental observations.

In a study similar to the present work, Mu et al.~\cite{mu_phase_2022} report that Al strongly prefers to occupy octahedral Ga sites in both $\beta$- and $\gamma$-\ch{Ga2O3}, unlike the results reported here for Mg, where this preference is only present in the $\beta$ phase. While Mg shows no strong site preference in $\gamma$-\ch{Ga2O3}, its incorporation does reduce the enthalpy difference between the $\beta$ and $\gamma$ phases, and the diffusion of Al and Mg during growth collectively make the $\gamma$ phase more energetically accessible. This is consistent with experimental observations that show Al preferentially occupies octahedral sites while Ga tends to occupy tetrahedral sites. Given that the $\gamma$ phase contains a higher proportion of octahedral to tetrahedral sites (5:3), and Mg does not show a site preference, it is more likely to occupy an octahedral site while stabilizing the $\gamma$ phase. 

While the enthalpies of formation are discussed in this work, it is important to consider that free energy, rather than enthalpy alone, governs phase stability under real synthesis conditions. In particular, configurational entropy, which arises from the number of ways atoms can be arranged in a structure can play a significant role in stabilizing disordered or metastable phases at finite temperatures. The $\gamma$-\ch{Ga2O3} phase is inherently structurally disordered, characterized by a defective spinel lattice with partial occupancy of both tetrahedral and octahedral cation sites. When Mg is introduced into this already disordered framework, the number of possible configurations increases dramatically due to the variability in choosing Ga sites for substitution, resulting in a high configurational entropy. This entropy contribution ($S= k_{\text B}\ln(\Omega)$; where $\Omega$ is the number of distinct configurations) lowers the free energy of the $\gamma$ phase relative to the other more ordered phases. This implies that high-temperature growth conditions or post-growth annealing treatments could thermodynamically favor the formation or persistence of the Mg-doped $\gamma$ phase.

We note also that in this work, we have assumed that the $\gamma$ phase remains ``\ch{Ga2O3}-like'' -- namely, that the cation:oxygen ratio remains approximately 2:3, with all deviations arising due to the additional vacancies introduced to preserve charge balance under Mg incorporation. In reality, experimental characterization suggests that at higher Mg and Al incorporation, these films may be better described as spinel-structured solid solutions. Future work may explore the broader spectrum of disorder that this implies.

\section{Conclusion}
In this work, we investigated the effect of Mg substitution on the relative stability of \ch{Ga2O3} polymorphs, with a particular focus on the $\beta$- and $\gamma$ phases that have been experimentally observed in thin films grown on Mg-containing substrates. Our calculations indicate that Mg incorporation reduces the enthalpy differences between the different phases of \ch{Ga2O3}, thereby lowering the energetic barrier to forming metastable phases.
For $\beta$-\ch{Ga2O3}, Mg exhibits a strong preference for octahedral Ga sites, and this preference becomes increasingly significant at higher Mg incorporation, where octahedral substitution substantially reduces the enthalpy of formation relative to other sites. In contrast, Mg shows no clear site preference in $\gamma$-\ch{Ga2O3}, consistent with the structural disorder of this phase. %A direct comparison between the $\beta$- and $\gamma$ phases reveals that Mg substitution at tetrahedral sites greatly reduces their enthalpy difference, making the $\gamma$ phase more energetically competitive.
When considered alongside prior work, these results provide a mechanism for the experimentally observed stabilization of $\gamma$-\ch{Ga2O3} on Mg-containing substrates. More broadly, the results highlight cation substitution as a practical route to tailoring polymorph stability in \ch{Ga2O3} semiconductors.

\subsection*{Code and Data Availability Statement}
Input structures, preprocessing code, and simulation input files are freely available on GitHub at \url{https://github.com/ACME-group-CMU/Ga2O3_Mg_paper}.

%%%%%%%%%%%%%%%%%%%%%%%%%%%%%%%%%%%%%%%%%%%%%%%%%%%%%%%%%%%%%%%%%%%%%
%% The "Acknowledgement" section can be given in all manuscript
%% classes.  This should be given within the "acknowledgement"
%% environment, which will make the correct section or running title.
%%%%%%%%%%%%%%%%%%%%%%%%%%%%%%%%%%%%%%%%%%%%%%%%%%%%%%%%%%%%%%%%%%%%%
\begin{acknowledgement}

% Please use ``The authors thank \ldots'' rather than ``The
% authors would like to thank \ldots''.

This research was conducted using the Tartan Research Advanced Computing Environment (TRACE). The authors would like to gratefully acknowledge the College of Engineering at Carnegie Mellon University for making this shared high-performance computing resource available to its community. The authors acknowledge use of the Materials Characterization Facility at Carnegie Mellon University supported by Grant No. MCF-677785, and funding from the National Science Foundation (Program Manager: Yaroslav Koshka, Grant No. DMR-2324375).

\end{acknowledgement}

%%%%%%%%%%%%%%%%%%%%%%%%%%%%%%%%%%%%%%%%%%%%%%%%%%%%%%%%%%%%%%%%%%%%%
%% The same is true for Supporting Information, which should use the
%% suppinfo environment.
%%%%%%%%%%%%%%%%%%%%%%%%%%%%%%%%%%%%%%%%%%%%%%%%%%%%%%%%%%%%%%%%%%%%%
\begin{suppinfo}

More detailed description of the process for selecting representative sets of alloy structures and associated figures.

\end{suppinfo}

%%%%%%%%%%%%%%%%%%%%%%%%%%%%%%%%%%%%%%%%%%%%%%%%%%%%%%%%%%%%%%%%%%%%%
%% The appropriate \bibliography command should be placed here.
%% Notice that the class file automatically sets \bibliographystyle
%% and also names the section correctly.
%%%%%%%%%%%%%%%%%%%%%%%%%%%%%%%%%%%%%%%%%%%%%%%%%%%%%%%%%%%%%%%%%%%%%
\bibliography{references}

\end{document}